\documentclass[aps,prd,twocolumn,showpacs,amsmath,amssymb]{revtex4-1}
\usepackage{graphicx}
\usepackage{subfig}
\usepackage{caption}
\usepackage{epstopdf}
\usepackage{color}
\usepackage{multirow}
\usepackage{setspace}
\usepackage{overpic}
\usepackage{amsmath}
\usepackage{amssymb}
\usepackage[bookmarksnumbered, pdfstartview=FitH,colorlinks,urlcolor=blue, citecolor=blue,linkcolor=blue] {hyperref}
\usepackage{lineno}
\usepackage{bm}
\usepackage{rotating}
\usepackage{verbatim}
\usepackage[utf8]{inputenc}
\graphicspath{{Fig/}} 
\usepackage{bigstrut}

\makeatletter
\renewcommand\tagform@[1]{\maketag@@@{(\ignorespaces#1\unskip\@@italiccorr)}}
\makeatother
\renewcommand{\eqref}[1]{(\ref{#1})}

\begin{document}


\title{\boldmath  Search for the lepton number violation decay $\omega \to \pi^+ \pi^+ e^-e^- +c.c.$ }


\author{
M.~Ablikim$^{1}$, M.~N.~Achasov$^{4,c}$, P.~Adlarson$^{77}$, X.~C.~Ai$^{82}$, R.~Aliberti$^{36}$, A.~Amoroso$^{76A,76C}$, Q.~An$^{73,59,a}$, Y.~Bai$^{58}$, O.~Bakina$^{37}$, Y.~Ban$^{47,h}$
% ...
\\
\vspace{0.2cm}
(BESIII Collaboration)\\
\vspace{0.2cm} {\it
$^{1}$ Institute of High Energy Physics, Beijing 100049, People's Republic of China\\
% ...
$^{4}$ Budker Institute of Nuclear Physics SB RAS (BINP), Novosibirsk 630090, Russia\\
% ...
$^{36}$ Johannes Gutenberg University of Mainz, Johann-Joachim-Becher-Weg 45, D-55099 Mainz, Germany\\
$^{37}$ Joint Institute for Nuclear Research, 141980 Dubna, Moscow region, Russia\\
% ...
$^{47}$ Peking University, Beijing 100871, People's Republic of China\\
% ...
$^{58}$ Southeast University, Nanjing 211100, People's Republic of China\\
$^{59}$ State Key Laboratory of Particle Detection and Electronics, Beijing 100049, Hefei 230026, People's Republic of China\\
% ...
$^{73}$ University of Science and Technology of China, Hefei 230026, People's Republic of China\\
% ...
$^{76}$ University of Turin and INFN, (A)University of Turin, I-10125, Turin, Italy; (B)University of Eastern Piedmont, I-15121, Alessandria, Italy; (C)INFN, I-10125, Turin, Italy\\
$^{77}$ Uppsala University, Box 516, SE-75120 Uppsala, Sweden\\
% ...
$^{82}$ Zhengzhou University, Zhengzhou 450001, People's Republic of China\\
\vspace{0.2cm}
$^{a}$ Deceased\\
% ...
$^{c}$ Also at the Novosibirsk State University, Novosibirsk, 630090, Russia\\
% ...
$^{h}$ Also at State Key Laboratory of Nuclear Physics and Technology, Peking University, Beijing 100871, People's Republic of China\\
% ...
}
}

\begin{abstract} 
The lepton number violation decay $\omega \to \pi^+ \pi^+ e^-e^-
+c.c.$ is searched for via $J/\psi \to \omega\eta$ using a data sample
of $(1.0087 \pm 0.0044) \times 10^{10}$ $J/\psi$ events collected by
the BESIII detector at the BEPCII collider. No significant signal is
observed, and the upper limit on the branching fraction of $\omega \to
\pi^+ \pi^+ e^-e^- +c.c.$ at the 90\% confidence level is determined
for the first time to be $2.8 \times 10^{-6}$.
\end{abstract}

\maketitle

\section{Introduction}

Neutrinos, described by the Dirac equation and considered as $\rm
SU(2)_{\rm L}$ gauge invariant fields, were accepted in the Standard
Model (SM) as massless left-handed Dirac fermions after the
experimental measurement of neutrino helicity as -1 in
1958~\cite{Helicity}. However, the Solar Neutrino Experiment, Sudbury
Neutrino Observatory, and Super-Kamioka Neutrino Detection Experiment
have observed neutrino oscillation~\cite{Super-Kamiokande
Collaboration, SNO Collaboration, KamLAND Collaboration, DAYA-BAY
Collaboration}, indicating that neutrinos possess a tiny mass.

If the antiparticle of a neutrino is itself, the solution to the Dirac
equation can give rise to a Majorana particle. In the theory of
Majorana~\cite{Majorana}, neutrinos can potentially possess mass. Some
theories, such as the seesaw mechanism~\cite{seesaw1,seesaw2}, provide
a natural framework for generating a small Majorana mass. If neutrinos
are indeed Majorana fermions, it would lead to a violation of lepton
number conservation by two units. Hence, the discovery of lepton
number violating processes might be relevant to the properties of
neutrinos. Furthermore, Baryon Number Violation (BNV) is a key aspect
of some Grand Unified Theories (GUTs) and is essential for
understanding the early Universe. The connection between BNV and
Lepton Number Violation (LNV) in various theories and
models~\cite{Fukugita M,Khlebnikov,Leptogenesis,sterileneutino}
suggests the search for the LNV decay is one of the important
approaches to establish a theory beyond the SM.

Various LNV signals have been sought after, and the search for
neutrinoless double-beta ($0\nu \beta \beta$)
decay~\cite{NeuDouBetaE1,NeuDouBetaE2,NeuDouBetaE3,NeuDouBetaE4},
which was first proposed by Furry~\cite{W. H. Furry} in 1939, is
considered to be the most sensitive.  Many collider experiments, such
as LHCb~\cite{LHCb}, CMS~\cite{CMS}, BaBar~\cite{Babar},
ATLAS~\cite{ATLAS}, CLEO~\cite{CLEO}, FOCUS~\cite{FOCUS},
BESIII~\cite{BESIII,rarecharm}, searched for the LNV decay.
E865~\cite{E865}, NA62~\cite{NA62}, and BESIII~\cite{phi} experiments
also searched for LNV with non-first generation quark decays in $K$
and $\phi$ meson decays, but so far have reported negative results
only.

\begin{figure}[htb]
	\centering
			\includegraphics[width=0.7\linewidth]{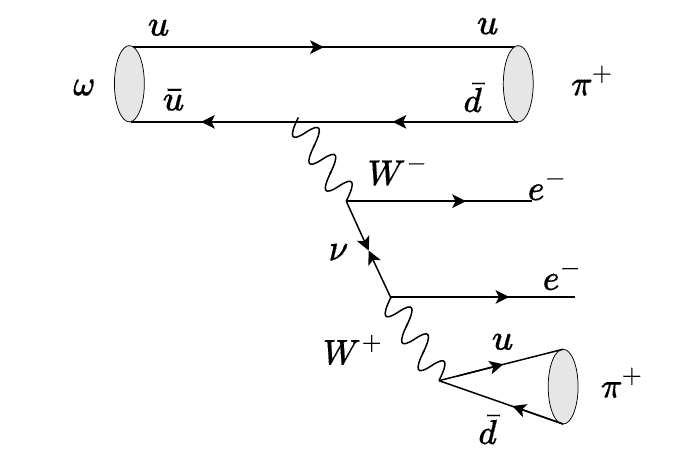}
	\caption{A possible Feynman diagram for $\omega \to \pi^{+}
	\pi^{+} e^{-} e^{-}$.}
	\label{F1}
\end{figure}

The world's largest $J/\psi$ dataset taken at BESIII offers a good
opportunity to search for possible LNV decays of various light
hadrons, e.g., $\omega\to \pi^+\pi^+e^-e^-$. Figure \ref{F1} shows a
possible Feynman diagram for $\omega \to \pi^+ \pi^+ e^- e^- $ in the
Majorana neutrino scenario, which would be suppressed by the $W^{\pm}$
bosons. The LNV decay of the $\omega$ meson in the process $\omega\to
\pi^+\pi^+e^-e^-$ has a unique phase space coverage compared to other
measurements and low background contamination.  Its discovery will
indicate the existence of new physics.

In this paper, we present the first search for the LNV decay $\omega
\to \pi^+ \pi^+ e^- e^-$ via $J/\psi\to\omega\eta$ decay based on
$(1.0087\pm 0.0044)\times 10^{10}$ $J/\psi$ events~\cite{Jpsidata}
collected by the BESIII~\cite{BESIIIDesign} detector at the Beijing
Electron-Positron Collider II (BEPCII)~\cite{BEPCII}.  The
charge-conjugated decay mode $\omega \to \pi^- \pi^- e^+ e^+ $ is
included and implicitly assumed throughout.

\section{BESIII detector and Monte Carlo simulation} 

BESIII~\cite{BESIIIDesign} is a symmetric cylindrical
particle detector located around the interaction point of
BEPCII~\cite{BEPCII}, which is an $e^{+}e^{-}$ collider employing a
double storage ring. The center-of-mass
collision luminosity of BEPCII reached a peak of $\rm 1.1\times
10^{33} \ cm^{-2}s^{-1}$ at $3.773~\rm GeV$. The BESIII detector
consists of four detector sub-components~\cite{BESIIIDesign}: a
helium-based multilayer drift chamber (MDC), a
plastic scintillator Time-Of-Flight counter (TOF), a CsI(Tl)
Electromagnetic Calorimeter (EMC), and a muon counter, providing a
coverage of 93\% of the total solid angle. The superconducting
solenoid supported by an octagonal flux-return yoke provides a
magnetic field of 1.0 T for the MDC for most of the $J/\psi$
data. The magnetic field was 0.9 T in 2012, which affects 11\% of the
total $J/\psi$ data.  The momentum resolution of the MDC for charged
particles at 1 GeV/\emph{c} is 0.5\%, and the ionization energy loss ($dE/dx$)
resolution for electrons from Bhabha scattering is 6\%. The time
resolution of the TOF barrel region is 68 ps, and the time resolution
of the end cap region was 110 ps. The end cap TOF system was upgraded
in 2015 using multigap resistive plate chamber technology, providing a
time resolution of 60 ps~\cite{upgradeBEPCII}, which benefits 87\% of
the data used in this analysis. The EMC achieves an energy resolution
of 2.5\% at 1 GeV.

Monte Carlo (MC) samples are used to analyze backgrounds and determine
the detection efficiency. The detector response, geometric
description~\cite{BESIII2, BESIII3}, and the signal digitization
models are simulated by {\sc Geant4}~\cite{GEANT4} software. For the
inclusive MC sample, the known $J/\psi$ decay modes are generated by
{\sc evtgen}~\cite{EVTGEN} with average branching fractions taken from
the Particle Data Group (PDG)~\cite{PDG}, while the remaining unknown
decays modes from the charmonium states are generated by {\sc
  lundcharm}~\cite{LUNDCHARM}. The $J/\psi$ resonance is produced via
$e^+e^-$ annihilations by {\sc kkmc}~\cite{KKMC}, which includes the
effects of beam energy spread and initial state radiation. Final state
radiation from charged final-state particles is incorporated with {\sc
  photos}~\cite{PHOTOS}. Exclusive MC events of $J/\psi\to
\omega\eta$, with $\eta \to \gamma\gamma$ and $\omega \to
\pi^{+}\pi^{-}\pi^{0}$ or $\omega\to\pi^{+}\pi^{+} e^{-}e^{-}$, are
also generated. The $J/\psi\to \omega\eta$ decays are modeled by a
helicity amplitude model~\cite{EVTGEN}, and the $\omega \to
\pi^{+}\pi^{-}\pi^{0}$ decays are modeled by an $\omega$ Dalitz
model~\cite{omegadalitz}. Other decays are modeled by a phase space
model.

\section{Data Analysis} 
%
%

To avoid the large uncertainty (11.5\%)~\cite{PDG} from the world
average value of $\mathcal{B}(J/\psi\to\omega\eta)$, we measure the
branching fraction of the decay $\omega\to\pi^+\pi^+e^-e^-$ relative
to the reference decay $\omega\to\pi^+\pi^-\pi^0$
\begin{linenomath*}
\begin{equation}\label{brachingfraction}
\begin{split}
\mathcal{B}(\omega\to\pi^+\pi^+e^-e^-)&=\mathcal{B}(\omega\to\pi^{+}\pi^{-}\pi^{0})\\
\times\mathcal{B}(\pi^{0}\to\gamma\gamma)&\times\frac{N^{\rm sig}_{\pi^+\pi^+e^-e^-}/\epsilon_{\pi^+\pi^+e^-e^-}}{N^{\rm ref}_{\pi^+\pi^-\pi^0}/\epsilon_{\pi^+\pi^-\pi^0}},
\end{split}
\end{equation}
\end{linenomath*}
where $\mathcal{B}(\omega\to\pi^{+}\pi^{-}\pi^{0})$ and
$\mathcal{B}(\pi^{0}\to\gamma\gamma)$ are the branching fractions of
$\omega\to\pi^{+}\pi^{-}\pi^{0}$ and $\pi^{0}\to\gamma\gamma$,
respectively. $N^{\rm sig}_{\pi^+\pi^+e^-e^-}$ and $N^{\rm
  ref}_{\pi^+\pi^-\pi^0}$ are the numbers of signal and reference channel events, respectively. $\epsilon_{\pi^+\pi^+e^-e^-}$ and $\epsilon_{\pi^+\pi^-\pi^0}$ are the
detection efficiencies of signal and reference decays, respectively.

%
%

All charged tracks are reconstructed in the detector acceptance region
of the MDC. Their polar angles $\theta$ are required to satisfy
$\left| \cos\theta \right| < 0.93$, where $\theta$ is measured
relative to the $z$-axis, the symmetry axis of the MDC. The distances
of closest approach to the interaction point of the charged tracks,
along the $z$ direction and in the plane perpendicular to the $z$-axis,
$\left|V_z\right|$ and $\left|V_{xy}\right|$, are required to be less
than 10 cm and 1 cm, respectively.

%
%
For charged particle identification (PID), we make use of a
combination of $dE/dx$ in the MDC, the time of flight in the TOF and
the information of clusters in the EMC to calculate the confidence
level (CL) for the pion, kaon and electron hypotheses (${\rm
  CL}_{\pi}$, ${\rm CL}_K$, ${\rm CL}_e$).  Pion candidates are
required to satisfy ${\rm CL}_{\pi}>0.001$ and ${\rm CL}_{\pi}>{\rm
  CL}_K$, while electron candidates are required to satisfy ${\rm
  CL}_e>0.001$ and ${\rm CL}_e/({\rm CL}_e+{\rm CL}_K+{\rm
  CL}_{\pi})>0.8$.

%
%
Photons are reconstructed using isolated clusters in the EMC. The
deposited energies in the barrel region ($\lvert \cos\theta
\rvert<0.8$) and endcap region ($0.86<\lvert \cos\theta \rvert<0.92$)
are required to be larger than $25~\rm MeV$ and $50~\rm MeV$,
respectively. To suppress electronic noise and unrelated encodings, the
EMC timing of the photon candidate is required to be within 700 ns
after the event start time. To eliminate photons emanating from charged
tracks, the opening angle between the photon and the nearest charged
track is required to be larger than 10 degrees.

\subsection{Analysis of \boldmath{$\omega\to\pi^{+}\pi^{-}\pi^{0}$}} 

For the reference channel $J/\psi\to\omega\eta$, with
$\eta\to\gamma\gamma$, $\omega\to\pi^{+}\pi^{-}\pi^{0}$ and
$\pi^{0}\to\gamma\gamma$, at least four reconstructed photons and two
reconstructed charged tracks with zero net charge are required.  To
select the photons of the $\pi^0$ and $\eta$ candidates, we calculate
the value of $\Delta m_{2\gamma}^{2} =
\frac{(M_{\gamma\gamma}-M_{\pi^{0}})^{2}}{\sigma_{\pi^{0}}^{2}}+\frac{(M_{\gamma\gamma}-M_{\eta})^{2}}{\sigma_{\eta}^{2}}$
for all possible sets of four photons, where $M_{\gamma \gamma}$ is
the $\gamma \gamma$ invariant mass, $M_{\pi^0}$ and $M_{\eta}$ are the
  known $\pi^0$ and $\eta$ masses~\cite{PDG}, and $\sigma_{\pi^{0}}$ and
  $\sigma_{\eta}$ are the corresponding mass resolutions determined
  from MC simulation. The $\pi^0$ and $\eta$ candidates with the
  smallest $\Delta m_{2\gamma}^{2}$ are kept for further analysis.

In order to reduce backgrounds and improve the mass resolution, a
five-constraint (5C) kinematic fit~\cite{KIMFIT} is performed to all
the tracks enforcing energy and momentum conservation and constraining
$M_{\gamma\gamma}$ to $M_{\pi^0}$.  The $\chi^2_{\rm 5C}$ from the
kinematic fit is required to be less than 20, which is determined by
optimizing the figure-of-merit $\frac{S}{\sqrt{S+B}}$, where $S$ is
the number of signal events and $B$ is the number of background events
from the inclusive MC sample. The requirement vetoes 84\% of
background contributions and retains 67\% of the reference
channel.

Backgrounds are investigated using the $1.0011\times 10^{10}$
inclusive $J/\psi$ MC events. Backgrounds with peaks in the invariant
mass distributions of both $\pi^+\pi^-\pi^0$ ($M_{3\pi}$) and
$M_{\gamma\gamma}$ are negligible.  The remaining
backgrounds includes non-peaking backgrounds (BKGI) and those with
peaks in either the $M_{3\pi}$ or $M_{\gamma\gamma}$ distributions (BKGII).
The contributions of these backgrounds are determined by performing a
two-dimensional (2D) fit to the invariant mass distributions of
$M_{3\pi}$ and $M_{\gamma\gamma}$.

\begin{figure}[hbt]
\centering
  \subfloat[]{
    \begin{minipage}[t]{0.88\linewidth}
      \includegraphics[width=\linewidth]{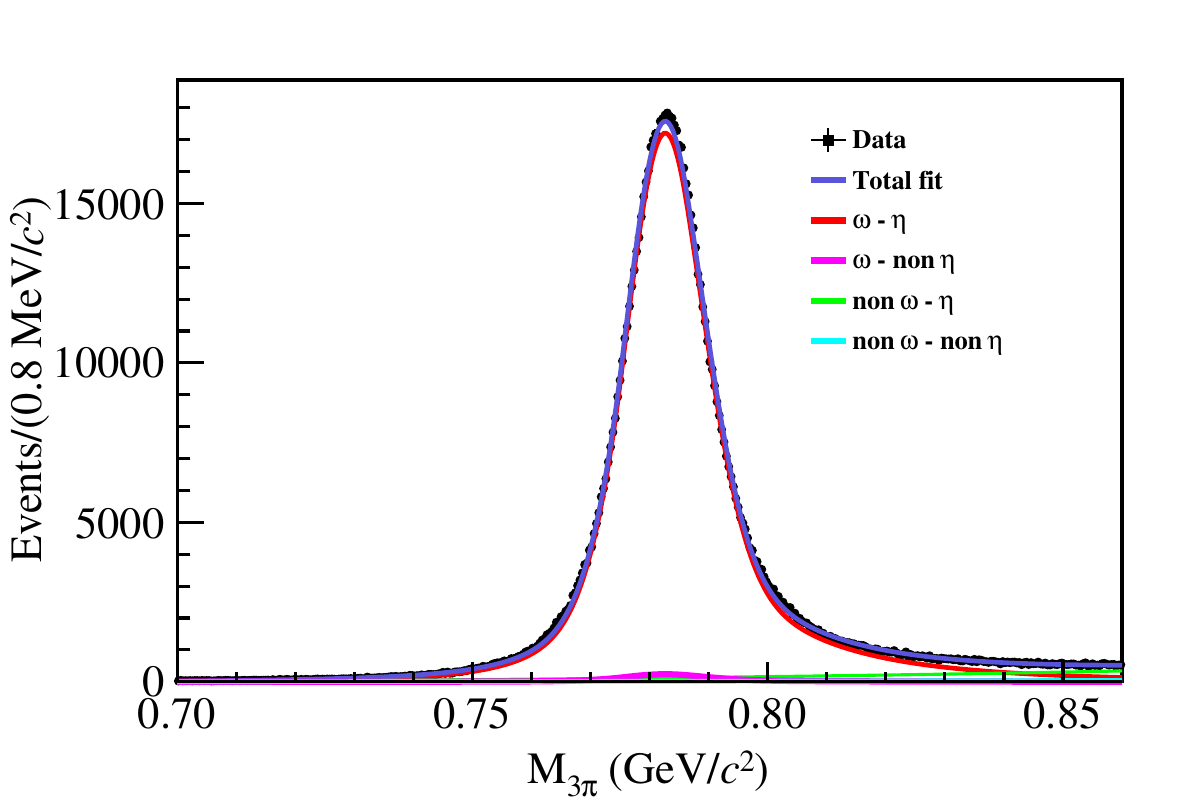}
    \end{minipage}
  }\\
  \subfloat[]{
    \begin{minipage}[t]{0.88\linewidth}
      \includegraphics[width=\linewidth]{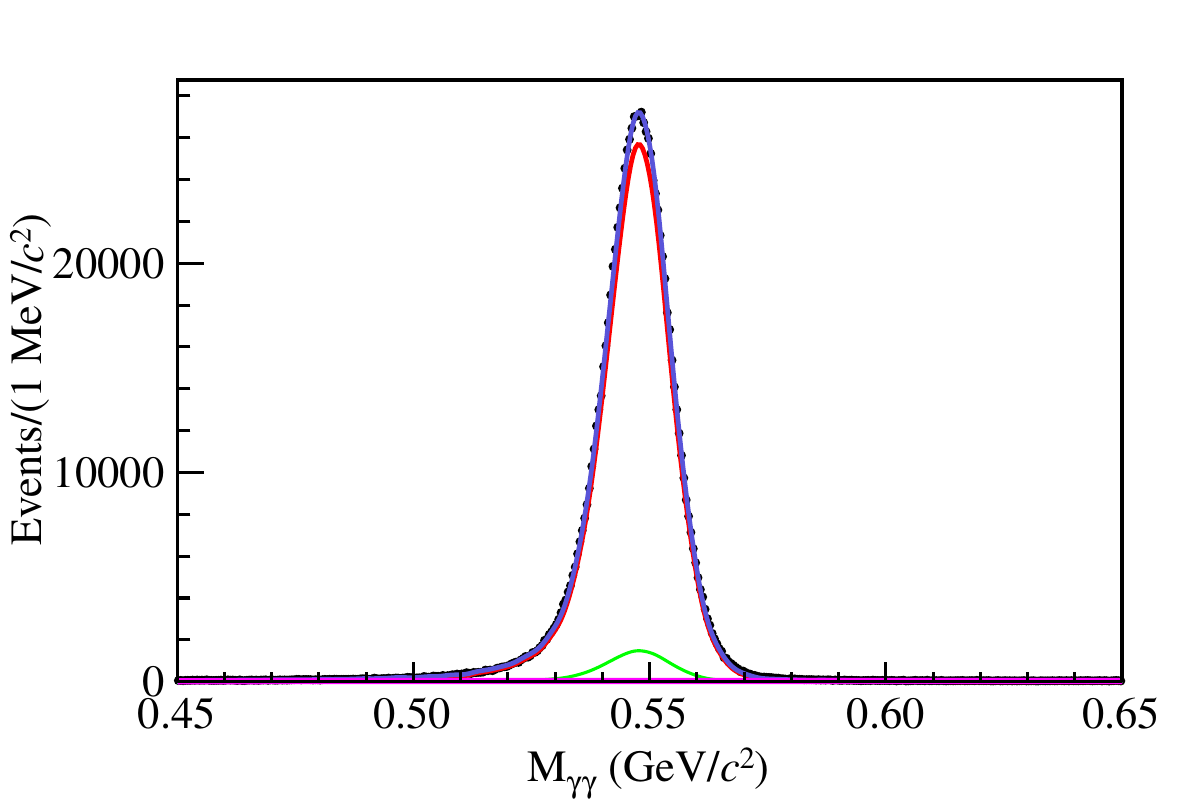}
    \end{minipage}
  }
\caption{The projections of the 2D fit of the (a) $M_{3\pi}$ and (b)
  $M_{\gamma\gamma}$ distributions, where the total PDF is shown by a blue
  solid curve. The signal PDF is shown by a red solid curve and
  labeled by $\omega - \eta$. The non-peaking background distribution
  is shown by a cyan solid curve and labeled by non-$\omega$ --
  non-$\eta$. The peaking background in $M_{3\pi}$ is shown by a pink
  solid curve and labeled by $\omega$ -- non-$\eta$. The peaking
  background in $M_{\gamma\gamma}$ is shown by a green solid line and
  labeled by non-$\omega$ -- $\eta$. } \label{fitm3pidata}
\end{figure}

We use the sum of the two Crystal Ball (CB) functions ($F_{\rm
  sig}^{\omega}$) with the same $\sigma$ and $\mu$ values but
different tail parameters to describe
the signal of $M_{3\pi}$ and a signal MC shape convolved with a
Gaussian function ($F_{\rm sig}^{\eta}$) to describe
$M_{\gamma\gamma}$. The non-peaking background contributions in the
$M_{3\pi}$  and $M_{\gamma\gamma}$ distributions are described by a
reversed ARGUS function~\cite{RA} ($F_{\rm bkg}^{\omega}$ and $F_{\rm
  bkg}^{\eta}$).  Consequently, the total signal shape is described by
$F_{\rm sig}^{\omega}\otimes F_{\rm sig}^{\eta}$, the backgrounds like
$\pi^+\pi^-\pi^0\eta$ and $\pi^0\pi^0\omega$ (BKGII) are described by
$F_{\rm bkg}^{\omega}\otimes F_{\rm sig}^{\eta}$ and $F_{\rm
  bkg}^{\eta}\otimes F_{\rm sig}^{\omega}$, and the non-peaking
background (BKGI) is described by $F_{\rm bkg}^{\omega}\otimes F_{\rm
  bkg}^{\eta}$.The fit range is chosen to be $[0.70,~0.86]
{\rm~GeV/\emph{c}^2}$ for $M_{3\pi}$, and $[0.45,~0.65]
{\rm~GeV/\emph{c}^2}$ for $M_{\gamma\gamma}$. We float the parameters of
probability density functions (PDFs) during the fit and the signal
yield is determined to be $N^{\rm ref}_{\pi^+\pi^-\pi^0}=941,336\pm
1,352$, where the uncertainty is statistical. The projections of the
2D fit to the $M_{3\pi}$ and $M_{\gamma\gamma}$ distributions are shown in
Fig.~\ref{fitm3pidata}.
The detection efficiency of the reference channel is determined by the
dedicated MC sample to be
$\epsilon_{\pi^+\pi^-\pi^0}=(12.80\pm0.03)\%$.  By taking into account
the signal yield $N^{\rm ref}_{\pi^+\pi^-\pi^0}$, the detection
efficiency $\epsilon_{\pi^+\pi^-\pi^0}$, and the decay branching
fractions of $\omega\to\pi^+\pi^-\pi^0$ and $\pi^0\to\gamma\gamma$
from the PDG~\cite{PDG}, the branching fraction of
$J/\psi\to\omega\eta$ is measured to be consistent with its world
average value quoted from the PDG~\cite{PDG} within one standard
deviation.

\subsection{Analysis of \boldmath{$\omega\to\pi^{+}\pi^{+}e^{-}e^{-}$}} 

For the signal channel $J/\psi\to\omega\eta$ with
$\eta\to\gamma\gamma$ and $\omega\to\pi^+\pi^+e^-e^-$, at least two
reconstructed photons and four reconstructed charged tracks with zero
net charge are required. The selection criteria of charged and neutral
tracks that are consistent with those used in the reference channel.

In order to reduce backgrounds and improve the mass resolution, a
four-constraint (4C) kinematic fit~\cite{KIMFIT} is performed by
constraining the energy and momentum of the four charged tracks
($\pi^+\pi^+e^-e^-$) and two photons to those of the initial state.
If there are more than two photons, we keep the candidate with the lowest
$\chi^{2}_{\rm 4C}$ for further analysis.  To further suppress
backgrounds, the $\chi^2_{\rm 4C}$ is required to be less than 10,
which is determined by the Punzi figure-of-merit method~\cite{Punzi},
with the figure of merit $\frac{\epsilon}{1.5+\sqrt{B}}$, where
$\epsilon$ is the detection efficiency and $B$ is the number of
background events.  This selection criterion can remove 99\% of
background events while retaining 56\% of the signal events.  To
further suppress misidentification from processes with four charged
tracks, we re-calculate the $\chi^{2}$ of the 4C kinematic fit with
different mass assignments ($\chi^{2}_{\rm re}$):
$\pi^+\pi^-\pi^+\pi^-\gamma\gamma$, $K^+K^-K^+K^-\gamma\gamma$,
$\pi^+\pi^-K^+K^-\gamma\gamma$, and $\pi^+\pi^-p
\bar{p}\gamma\gamma$. If any of the $\chi^{2}_{\rm re}$s is less than
$\chi^{2}_{\rm 4C}$, the event is considered as background and
rejected.

The signal region is determined by fitting the $M_{\pi^+\pi^+e^-e^-}$
and $M_{\gamma\gamma}$ distributions of signal MC samples, where the
signal shape is modeled by a double Gaussian function and the
background shape a second-order polynomial function.  The fitted
regions are $[0.72,0.84]~{\rm GeV}/c^2$ for $M_{\pi^+\pi^+e^-e^-}$ and
$[0.51,0.59]~{\rm GeV}/c^2$ for $M_{\gamma\gamma}$, which correspond
to $[\mu - 5\sigma , \mu + 5\sigma]$, where $\mu$ and $\sigma$ are the
fitted mean value and standard deviation.  The detection efficiency is
$\epsilon_{\pi^+\pi^+e^-e^-}=(11.52\pm0.03)\%$ based on
the simulated signal MC samples.

\begin{figure}[hbt]
	\centering
	\includegraphics[width=0.88\linewidth]{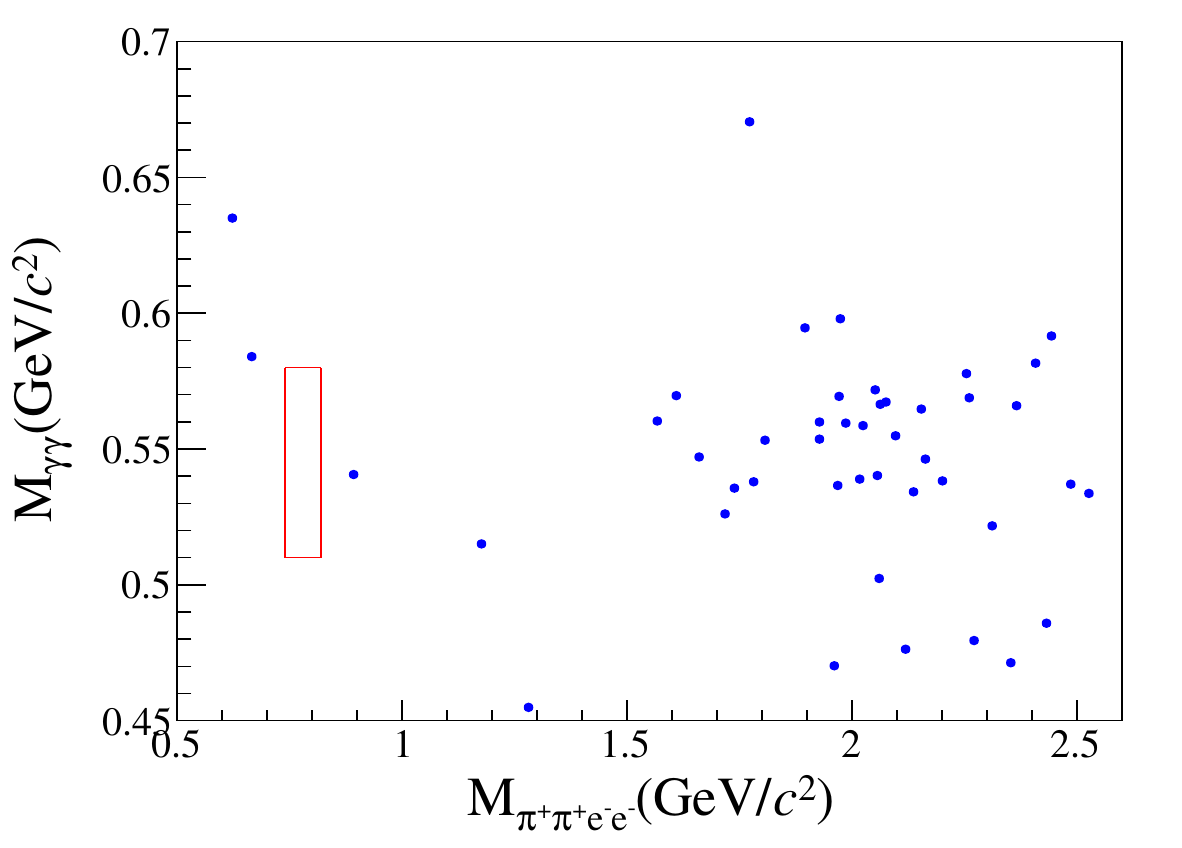}
	\caption{The 2D distribution of $M_{\gamma\gamma}$ versus
          $M_{\pi^+ \pi^+ e^-e^-}$ of the candidate events from the
          $J/\psi$ data, where the red box shows the signal region.}
	\label{metampipiee}
\end{figure}

Figure~\ref{metampipiee} shows the 2D scatter plot of
$M_{\gamma\gamma}$ versus $M_{\pi^+\pi^+e^-e^-}$ for the candidate
events from $J/\psi$ data. In the scatterplot no candidates are found
inside the defined signal region. Thus, the number of observed events
is $N^{\rm obs}_{\pi^+\pi^+e^-e^-}=0$.  To study the possible
backgrounds, $1.0011\times 10^{10}$ inclusive $J/\psi$ MC events are
analyzed. We find there is no event located near the signal region,
and the number of background events is determined to be $N^{\rm
  bkg}_{\pi^+\pi^+e^-e^-}=0$.

\subsection{Systematic uncertainty} 
The systematic uncertainties for the signal and reference decays
include the uncertainties in the MDC tracking, PID, $\pi^0$
reconstruction, kinematic fit and $\chi^{2}$ requirement, signal
window, 2D fit, MC modeling, $N^{\rm ref}_{\pi^+\pi^-\pi^0}$
determination, and input branching fractions. The systematic
uncertainties of the $\eta$ reconstruction with
$\mathcal{B}(J/\psi\to\omega\eta)$ and
$\mathcal{B}(\eta\to\gamma\gamma)$ cancel due to calculating the
relative branching fraction ratio.

The systematic uncertainties of MDC tracking and PID for charged pions
and electrons are 1.0\%~\cite{trackingPIDofelec,trackingPIDofpion} per
charged track. Since there are two charged tracks for the reference
channel and four charged tracks for the LNV channel, the systematic
uncertainties are assigned to be 2.0\% and 4.0\%, respectively. The
total systematic uncertainty for MDC tracking and PID of charged
tracks is calculated to be $\sqrt{(2.0\%)^2+(4.0\%)^2}=4.5\%$ for
each.

The systematic uncertainty of $\pi^0$ reconstruction is
1.0\%~\cite{etareconstruction}.  The systematic uncertainty in the
kinematic fit and the $\chi^{2}$ requirement of the signal (reference)
decay is estimated with the control sample
$J/\psi\rightarrow\pi^{+}\pi^{-}\pi^{+}\pi^{-}\eta,
\eta\rightarrow\gamma\gamma$ ($J/\psi\rightarrow\omega\pi^{0},
\omega\rightarrow\pi^{+}\pi^{-}\pi^{0}$).  The relative difference of
the selection efficiencies between data and MC simulation is taken as
the uncertainty. The systematic uncertainties due to the kinematic fit
are 2.7\% and 0.1\% for the LNV and reference decays,
respectively. The total systematic uncertainty is calculated to be
2.7\%.  The systematic uncertainty of signal window is studied by
changing the ranges to $\pm 4.9\sigma$, $\pm 5.1\sigma$, etc. The
maximum relative difference 0.2\% is taken as the systematic
uncertainty.

The uncertainties from the background (signal) shapes in the 2D fit
are estimated by changing the reverse ARGUS function (sum of two CB
function) to a second-order polynomial function (sum of a CB function
and the signal MC shape obtained from MC simulation). The maximum
differences of 3.3\% and 1.1\% between the signal yields are taken as
the uncertainties due to background and signal shapes. So the total
uncertainty for 2D fit is calculated to be 3.5\%.

The systematic uncertainty of the MC modeling is studied by generating
events with a Majorana intermediate state in two different modes:
one with two Majorana neutrinos ($\omega\rightarrow\nu_{M}\nu_{M},
\nu_{M}\rightarrow\pi^{+}e^{-}$), and the other with one Majorana
neutrino ($\omega\rightarrow\nu_{M}\pi^{+}e^{-},
\nu_{M}\rightarrow\pi^{+}e^{-}$), where the mass of the Majorana
neutrino can range from the $\pi e$ mass threshold to the largest
phase space of $\omega$ decay. The largest difference between the
average detection efficiencies in the nominal analysis and in these
two modes, 0.3\%, is taken as the systematic uncertainty.

The statistical uncertainty in $N^{\rm ref}_{\pi^+\pi^-\pi^0}$
determination is calculated to be 0.2\%.  The uncertainty due to MC
statistics is given by $\sqrt{\frac{1-\epsilon}{\epsilon N_{\rm
      total}^{\rm MC}}}$, where $\epsilon$ is detection efficiency of
$\omega\rightarrow\pi^+\pi^+e^-e^-$, and $N_{\rm total}^{\rm MC}$ is
the total number of produced signal MC events. It is
determined to be 0.3\%.  The uncertainty of the input branching
fraction $\mathcal{B}(\omega \to \pi^{+}\pi^{-}\pi^{0}, \pi^{0}\to
\gamma \gamma)$ is 0.8\%~\cite{PDG}.

Table~\ref{syssum} summarizes the systematic uncertainties. The
total systematic uncertainty ($\Delta_{\rm sys}$) is calculated by
adding the individual contributions in quadrature.

\begin{table}[hbt]
	\centering
	\caption{Relative systematic uncertainties.}
\begin{tabular}{lc}
	\hline
	\hline
	\multicolumn{1}{l}{Source} & Uncertainty~(\%)   \bigstrut\\
	\hline
	\multicolumn{1}{l}{MDC tracking}  & 4.5   \bigstrut[t]\\
	\multicolumn{1}{l}{PID}  &  4.5    \\
	\multicolumn{1}{l}{$\pi^0$ reconstruction}   & 1.0  \\
	\multicolumn{1}{l}{Kinematic fit and $\chi^{2}$ requirement }  & 2.7   \\
	\multicolumn{1}{l}{Signal window}    &  0.2    \\
	\multicolumn{1}{l}{2D fit}  &  3.5   \\
	\multicolumn{1}{l}{MC modeling} &  0.3    \\
	\multicolumn{1}{l}{$N^{\rm ref}_{\pi^+\pi^-\pi^0}$ determination} &  0.2   \\
	\multicolumn{1}{l}{MC statistics}   & 0.3    \\
	\multicolumn{1}{l}{$\mathcal{B}(\omega \to \pi^{+}\pi^{-}\pi^{0}, \pi^{0}\to \gamma \gamma)$} &   0.8    \bigstrut[b]\\
	\hline
	Total  & 7.9   \bigstrut\\
	\hline
	\hline
\end{tabular}%
	\label{syssum}%
\end{table}%

\section{Result} 
Since we do not observe any signal or background events, we
set an upper limit on the signal yield at the 90\% confidence level
(C.L.) using Feldman-Cousins intervals~\cite{Feldman}. Both the number
of observed events and the background yield are assumed to follow
Poisson distributions, and the upper limit on the signal yield is
calculated to be $2.44$ at the 90\% C.L. Since the Feldman-Cousins
method does not take into account the systematic uncertainty
($\Delta_{\mathrm{sys}}=7.9\%$), the upper limit 
is shifted up to
$2.44/(1-\Delta_{\mathrm{sys}})=2.65$. Thus, the upper limit on the
branching fraction of $\omega\to \pi^+\pi^+e^-e^-$ at the $90\%$
C.L. is calculated by Eq. \eqref{brachingfraction} to be
\begin{linenomath*}
\begin{equation*}
	\mathcal{B}(\omega\to \pi^+\pi^+e^-e^-)<2.8 \times 10^{-6}.
\end{equation*}
\end{linenomath*}

\section{Summary} 
In this paper, we search for the LNV decay $\omega \to \pi^{+} \pi^{+}
e^{-} e^{-}$ for the first time by analyzing $1.0087\times10^{10}$
$J/\psi$ events collected with the BESIII detector.  No signal is
observed, and the upper limit on its decay branching fraction is set
to be $\mathcal{B}(\omega \to \pi^+\pi^+e^-e^-+c.c.)<2.8\times10^{-6}$
at the 90\% C.L.  This is the first experimental constraint on the LNV
decay of the $\omega$ meson.

\section{Acknowledgements} 
The BESIII Collaboration thanks the staff of BEPCII (https://cstr.cn/31109.02.BEPC) and the IHEP computing center for their strong support. This work is supported in part by National Key R\&D Program of China under Contracts Nos. 2023YFA1606000, 2020YFA0406400, 2020YFA0406300, 2023YFA1606704; National Natural Science Foundation of China (NSFC) under Contracts Nos. 12035009, 11875170, 11635010, 11935015, 11935016, 11935018, 12025502, 12035013, 12061131003, 12192260, 12192261, 12192262, 12192263, 12192264, 12192265, 12221005, 12225509, 12235017, 12361141819; the Chinese Academy of Sciences (CAS) Large-Scale Scientific Facility Program; CAS under Contract No. YSBR-101; 100 Talents Program of CAS; The Institute of Nuclear and Particle Physics (INPAC) and Shanghai Key Laboratory for Particle Physics and Cosmology; Agencia Nacional de Investigación y Desarrollo de Chile (ANID), Chile under Contract No. ANID PIA/APOYO AFB230003; German Research Foundation DFG under Contract No. FOR5327; Istituto Nazionale di Fisica Nucleare, Italy; Knut and Alice Wallenberg Foundation under Contracts Nos. 2021.0174, 2021.0299; Ministry of Development of Turkey under Contract No. DPT2006K-120470; National Research Foundation of Korea under Contract No. NRF-2022R1A2C1092335; National Science and Technology fund of Mongolia; National Science Research and Innovation Fund (NSRF) via the Program Management Unit for Human Resources \& Institutional Development, Research and Innovation of Thailand under Contract No. B50G670107; Polish National Science Centre under Contract No. 2024/53/B/ST2/00975; Swedish Research Council under Contract No. 2019.04595; U. S. Department of Energy under Contract No. DE-FG02-05ER41374.

\end{document}